\documentclass[conference]{IEEEtran} 

\IEEEoverridecommandlockouts
\usepackage{graphicx}
\usepackage{epsfig}
\usepackage{subfigure}
\usepackage{tabularx}
\usepackage{amssymb}
\usepackage{latexsym}
\usepackage{amsmath}
\DeclareMathOperator*{\argmax}{arg\,max}

\usepackage{multirow}
\usepackage{algorithm}
\usepackage[noend]{algpseudocode}
\usepackage{color}
\usepackage{relsize}
\usepackage{mathtools}
\usepackage{amsthm}
\usepackage[,skip=3pt]{caption}
\usepackage{cite}

\usepackage{pifont}
\newcommand\Mycomb[2][^n]{\prescript{#1\mkern-0.5mu}{}C_{#2}}

\theoremstyle{definition}
\begin{document}

\title{A Practical AoI Scheduler in IoT Networks with Relays}

%
%
%


\author{Biplav Choudhury~\IEEEmembership{Student Member,~IEEE},\thanks{
Biplav Choudhury and Jeffrey H. Reed are with ECE Department at Virginia Tech. Prasenjit Karmakar is with CSE Department at IIT Kharagpur, and Vijay K. Shah is with Cybersecurity Engineering Department at George Mason University. 
(emails:\{biplavc, reedjh\}@vt.edu, prasenjitkarmakar52282@gmail.com and vshah22@gmu.edu).
This research was supported in part by ONR under MURI Grant N00014-19-1-2621 and by Virginia Commonwealth Cyber Initiative (CCI). CCI is an investment in the advancement of cyber R\&D, innovation, and workforce development (www.cyberinitiative.org). 
}
        Prasenjit Karmakar, 
        Vijay K. Shah~\IEEEmembership{Member,~IEEE}, \\
        and
        Jeffrey H. Reed~\IEEEmembership{Fellow,~IEEE}
}

\markboth{Journal of \LaTeX\ Class Files,~Vol.~14, No.~8, August~2021}%
{Shell \MakeLowercase{\textit{et al.}}: A Sample Article Using IEEEtran.cls for IEEE Journals}



\maketitle

\begin{abstract}



Internet of Things (IoT) networks have become ubiquitous as autonomous computing, communication and collaboration among devices become popular for accomplishing various tasks.  The use of relays in IoT networks further makes it convenient to deploy IoT networks as relays provide a host of benefits, like increasing the communication range and minimizing power consumption. Existing literature on traditional AoI schedulers for such two-hop relayed IoT networks are limited because they are designed assuming constant/non-changing channel conditions and known (usually, generate-at-will) packet generation patterns. Deep reinforcement learning (DRL) algorithms have been investigated for AoI scheduling in two-hop IoT networks with relays, however, they are only applicable for small-scale IoT networks due to exponential rise in action space as the networks become large. These limitations discourage the practical utilization of AoI schedulers for IoT network deployments. This paper presents a practical AoI scheduler for two-hop IoT networks with relays that addresses the above limitations. The proposed scheduler utilizes a novel voting mechanism based proximal policy optimization (v-PPO) algorithm that maintains a linear action space, enabling it be scale well with larger IoT networks. The proposed v-PPO based AoI scheduler adapts well to changing network conditions and accounts for unknown traffic generation patterns, making it practical for real-world IoT deployments. Simulation results show that the proposed v-PPO based AoI scheduler outperforms both ML and traditional (non-ML) AoI schedulers, such as, Deep Q Network (DQN)-based AoI Scheduler, Maximal Age First-Maximal Age Difference (MAF-MAD), MAF (Maximal Age First) , and round-robin in all considered practical scenarios.





\end{abstract}


\IEEEpeerreviewmaketitle

\section{Introduction}
\label{sec:intro}

 The Internet of Things (IoT) network has emerged as one of the most crucial technologies for providing connectivity. They have found widespread applications in healthcare, military, agriculture, industrial automation~\cite{atzori2010internet}, etc. and the total number of IoT devices all over the world is expected to exceed 30 billion by 2025~\cite{GlobalIo97:online}. While most IoT networks involve a direct connection between the IoT devices and a terrestrial base station (TBS) where the IoT devices send the data collected to the TBS for processing, recent works have also investigated the connection between the devices and the TBS being through a relay. E.g., single hop IoT networks are reported to suffer from connectivity issues due to lack of a direct line-of-sight (LOS) between the devices and the TBS~\cite{9540708}. In such cases, the relays provide a reliable means of forwarding the data collected by the IoT devices to the TBS. We refer to these two-hop IoT networks as \textit{IoT networks with relays}, or \textit{relayed IoT networks}, and they are extremely critical for deployment of many types of wireless services. For e.g., LoRaWAN is very critical for applications needing less throughput with long-range like smart grid, traffic monitoring etc, and relays play an important role in improving the range of LoRaWAN~\cite{8030482} which aids in its deployment. Another important advantage of using relays in IoT networks is that it improves power consumption~\cite{9912934}. Due to the these factors, the LoRa Alliance has officially announced a relay feature for LoRaWAN as an enabler for massive IoT networks~\cite{IoTNewsG83:online}.

In these IoT networks with relays, it is crucial that the information the TBS receives via the relays from the IoT devices is \textit{fresh}. This is desirable as fresher the information at the TBS, the better is its estimate on the real-time state of the IoT devices. The freshness of information is measured using the metric called ``Age of Information" (AoI) and was first introduced in~\cite{5984917}. A detailed survey for AoI-related works is presented in~\cite{9380899} and an online bibliography can be found in~\cite{webhomea5:online}. AoI has seen an active interest from the research community and has been investigated for general IoT networks, UAV networks, vehicular networks, and other time-sensitive applications~\cite{9151993,choudhury2020joint,9305697,9129400}. While the majority of research in AoI minimization in IoT networks deal with single-hop IoT networks without relays, there are some works that investigate AoI minimization in IoT networks with relays~\cite{9014214}. Refer to Related works section for details. Existing works on relayed IoT networks are limited in their practical utility for they suffer from the following shortcomings -

\begin{itemize}


    \item \textit{Accounting for unknown traffic generation patterns and channel qualities/conditions} - Traditional (non-ML) approaches cannot account for unknown traffic generation patterns and channel qualities. Most of theoretical study usually make assumptions of ideal channels and/or generate-at-will packet generation patterns, which is not the case in real-world IoT networks. There are some works~\cite{li2019kronos} that assume periodic or random packet generation models, however, they are assumed to be known apriori to the scheduler at the time of scheduling IoT devices, which is usually not available in real-world IoT networks. This area has not been explored much in literature and is a major bottleneck for practical IoT network deployments~\cite{9141208}

    \item \textit{Scalability to larger IoT networks} - Traditional queuing theory-based and optimization based schedulers are applicable to larger IoT networks, however, such IoT networks are idealistic, and suffer from the above limitation of unknown traffic generation patterns and channel qualities/conditions. Recently, deep reinforcement learning, particularly, Deep Q networks (DQN) based AoI schedulers has been investigated for designing AoI schedulers that can account for varying channel conditions. However, such ML-based approaches only work for smaller IoT networks (with 5-10 IoT devices). This is because the action space for DQN increases exponentially, which makes it impractical to utilize for scheduling in real-world IoT networks.
    
\end{itemize}


In this paper, we introduce a practical Age of Information (AoI) scheduler for two-hop IoT networks with relays, which overcomes the aforementioned limitations. Our proposed AoI scheduler employs a novel voting mechanism-based proximal policy optimization (v-PPO) algorithm that maintains a linear action space, making it capable of scaling well with larger IoT networks. Additionally, the v-PPO based AoI scheduler adapts well to changing network conditions and can handle unknown traffic generation patterns, even for large scale networks. We believe that our proposed v-PPO based AoI scheduler can be practically deployed for improving information freshness in real-world IoT networks with relays.



The main contributions of this paper are the following-

\begin{itemize}
    \item This paper proposes a practical AoI scheduling solution to address the key challenges faced in real-world IoT networks with relays, including, changes in the network conditions, unknown traffic generation patterns and channel qualities, and scalability to larger IoT networks. Our proposed AoI scheduler for relayed IoT networks overcomes these issues, and can be practically deployed for AoI-minimizing scheduling in real-world IoT networks.

    \item Our proposed scheduler utilizes a novel voting mechanism based proximal policy optimization (v-PPO) algorithm that maintains a linearly increasing action space with the increasing IoT network size. In other words, Our v-PPO based AoI scheduler scales well with large IoT networks. Furthermore, the proposed v-PPO based AoI scheduler also adapts well to the changing network conditions and learns well the unknown traffic generation patterns, making it a practical solution approach.
    
    

    \item Our simulation results show that the proposed v-PPO based AoI scheduler outperforms both traditional (non ML) and ML-based AoI schedulers, namely, Deep Q Networks (DQN) based scheduler, Maximal Age First-Maximal Age Difference (MAF-MAD), Maximal Age First (MAF), Round Robin (RR), and random schedulers under all considered practical simulation scenarios. 
\end{itemize}

The rest of the paper is as follows: Sec.~\ref{sec:related} discusses the related works and Sec.~\ref{sec:system} describes the system model. In Sec.~\ref{sec:ppo_policy}, we explain the proposed v-PPO-based AoI scheduler and the performance in minimizing AoI for all the schedulers is compared in Sec.~\ref{sec:results}. The paper is concluded in Sec.~\ref{sec:conclusion}.

\section{Related Work}
\label{sec:related}


The application of relays is being increasingly researched for IoT networks, and its utility in AoI minimization in IoT networks has seen significant interest from the research community. Generally, they can be clubbed into ML and non-ML approaches. Some of the non-ML works are - the AoI minimization in multi-hop energy harvesting wireless sensor networks is studied in~\cite{chen2022minimizing} based on the update generation time with the objective of minimizing the peak AoI and average AoI. Reference~\cite{9754994} studies the AoI of two-way relay networks operated with physical-layer network coding with and without automatic repeat request. In~\cite{buyukates2018age}, a single source node is considered that is transmitting to multiple receiver nodes via relays based on receiving an acknowledgment. As the AoI at the receiver node is dependent on the waiting time, the optimal waiting time is calculated. A single energy constrained source transmitting its information to a single receiver via an energy constrained relay is considered in~\cite{arafa2019timely}, and AoI minimizing online and offline policies are investigated. The authors in~\cite{moradian2020age} do a discrete-time stochastic hybrid system analysis for calculating the AoI with and without relays between a single transmitter and a single receiver. It can be seen that in most of the works mentioned above, the network conditions are presumed to be known which allows for it to be solved using classical optimization techniques. However, it is often the case that the exact network conditions are not known and also they might change with time. In such cases, such approaches will fail to perform.

Some of the ML based works include - reference~\cite{song2020optimal} considers a single relay which samples information from multiple IoT devices to their respective destinations. However,~\cite{song2020optimal} only considers a generate-at-will traffic model at the IoT devices with only a single relay. In \cite{9576539}, UAVs relay traffic from vehicles to a base station but it only considers a fixed traffic generation pattern at the vehicles based on a Poisson process. In~\cite{eldeeb2022multi}, the authors propose a scheme to jointly plan trajectory of UAVs acting as relays while minimizing the AoI and energy consumption for an IoT network. Our work is closest to~\cite{9637803} where the authors study AoI-minimizing for UAV networks with the UAVs as relays between IoT devices and TBS. However, the scenarios considered in the above either involve very small number of relays, or doesn't take into account changing network conditions, or both. This work deals with IoT networks with relays with a large number of relays and incorporates changing network conditions, which makes it generalizable and suitable for practical deployments.

\section{Network Model and Problem Formulation}
\label{sec:system}

\begin{figure} [htb]
    \centering
    \includegraphics[width=\columnwidth]{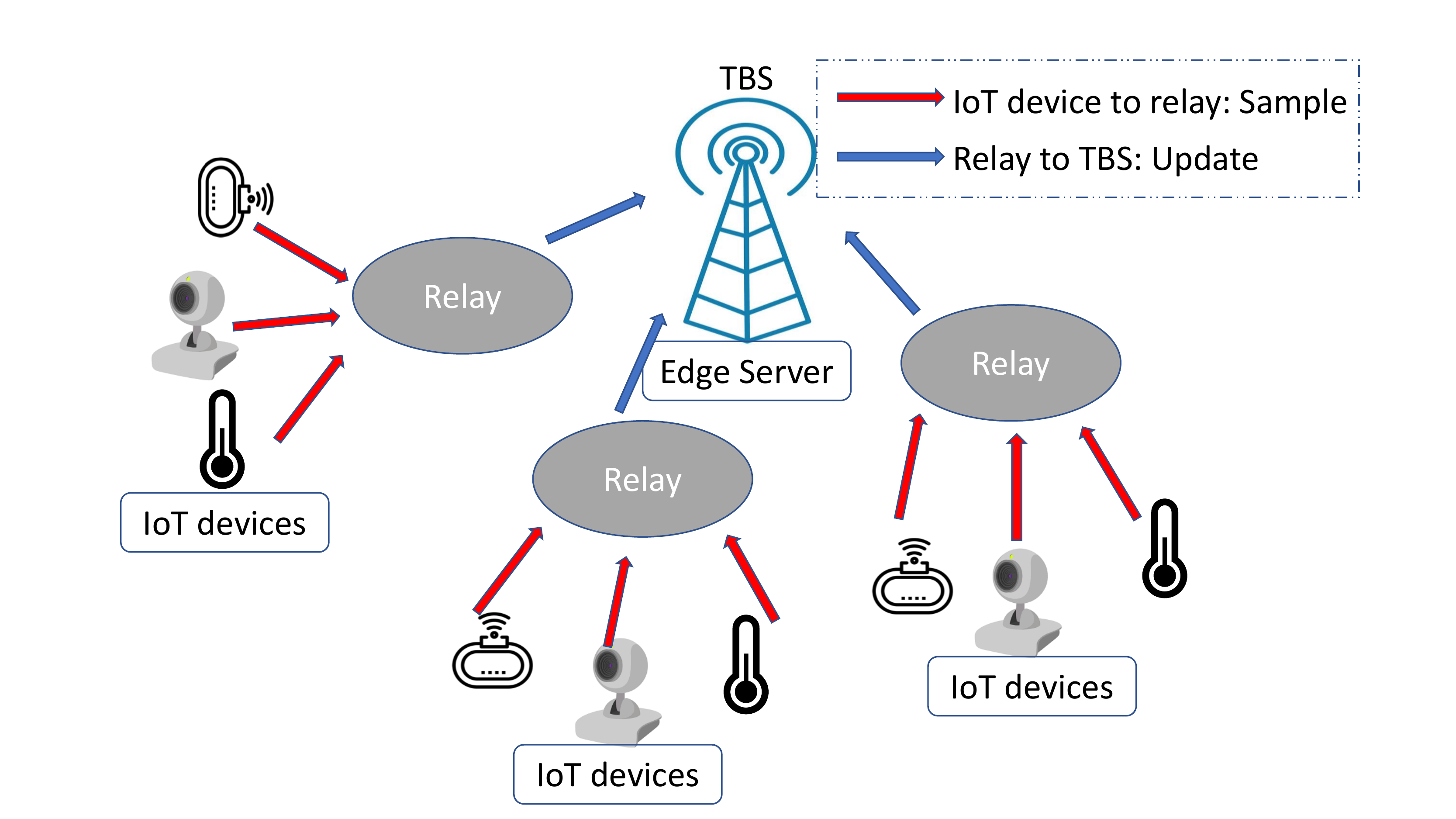}
    \caption{Overview IoT networks with relays}
    \label{fig:scenario}
    \vspace{-0.05in}
\end{figure}

\subsection{Network Model}


As shown in Fig.~\ref{fig:scenario}, a relayed IoT network comprises of a TBS, $M$ IoT devices and $N$ relays. The TBS is located at a fixed 2D ground location and it is equipped with an edge server to process the collected time-sensitive information. $M$ IoT devices are distributed randomly~\cite{lyu2016placement} in the considered region. We consider that the IoT devices are energy-constrained\footnote{Note that the energy consumption of the IoT devices and relays are outside the scope of the paper as the focus in on the scheduling aspect.} and have limited wireless range\footnote{Most IoT devices have a wireless range of less than $100$ meters~\cite{HowIoTSh99:online}.}, and are thus outside the wireless coverage of the nearby TBS, such as, in case of rural agricultural smart farms, where the IoT devices are far away from the nearest TBS. Therefore $N$ relays are deployed in the deployment area and act as \textit{communication relays} between the nearby TBS and all $M$ IoT devices. Each relay $n \in N$ is deployed in the region of interest such that  

(i) relay $n$ lies within the wireless coverage of the TBS and, provides \textit{backhaul links} between itself and TBS and 

(ii) relay $n$ provides wireless coverage as \textit{access links} to $m_n$ subset of IoT devices, where $m_n \subseteq M$. 

Next, we discuss different traffic generation models for each IoT device and the wireless channels considered in our model. Assume $T$ is the total observation duration and it is divided into equal-length time slots denoted by $t$. 

\textbf{Traffic generation at the IoT devices.} We consider two different types of traffic generation models at each IoT device $m$ -- (i) \textit{generate-at-will}: device $m$ generates a new packet whenever it is selected to sample~\cite{abd2019deep}, and (ii) \textit{ (Unknown) periodic packet generation}: device $m$ generates information packets at a fixed time interval denoted by its periodicity $p_m$. It means, $m$ generates a new packet at slots $p_m t, 2p_m t, \dots, \lfloor \frac{T}{p_m} \rfloor t$. Note that though $p_m$ is a fixed value for a certain IoT device $m$, it is unique to each IoT device, i.e., $p_m \neq p_{m'}$, where $m, m' \in M$. Also, the periodicity $p_m$ for any IoT device is unknown beforehand, and may change over time. 

While the generate-at-will model is extensively used in the AoI literature as it is easy to model, it is not realistic to assume that the IoT devices generate a new packet whenever it is sampled. In most IoT applications, the IoT devices sense a physical phenomenon and generate packets periodically~\cite{aziz2019effective} and hence this is one of the traffic models considered by 3GPP for IoT applications in environment monitoring and control~\cite{8985528}. Moreover, in certain scenarios, the devices may change their periodicities. For example, a device may start generating packets at a faster/slower rate in response to application demands at the TBS, or network functions and protocols~\cite{xi2019hypergen}.

\textbf{Channels:} Let $K$ denote the number of channels for communication between the TBS and all $N$ relays. Similarly, let $L_n$ denote the total number of channels available between a certain relay $n$ and it's associated IoT devices. While our model can be easily extended to unique $L_n$ channels for access links between a relay and its associated IoT devices, we consider a fixed $L = L_n, (\forall n \in N$) channels for access links between a certain relay and it's associated IoT devices for ease of presentation. Even though a certain IoT device may fall within the wireless coverage of two or more relays, we assume that each IoT device is associated to a unique relay based on an association policy (e.g., max power association policy~\cite{8723312}), and thus, the subset of IoT devices, $m_n$, is unique to each relay, and $M =\cup_{i=1}^{N} m_i$. 

Due to various factors like noise, interference or other factors, packets can get lost while being transmitted from the IoT devices to the relay, or from the relays to the TBS. We capture these link outages by considering that the link between device $m$ and the associated relay $n_m $ has a non-zero probability of dropping the transmitted packets~\cite{song2020optimal, 9637803,8885156}, which we refer to as, \textit{Sample Loss Probability} and is denoted by $l_{s,m} \in (0,1)$. Similarly, $l_{u,m} \in (0,1)$ denotes the \textit{Update Loss Probability}, which refers to the packet loss probability between relay $n_m$ and the TBS. While our approach can be used for any channel model, we pre-assign an expected loss probability for ease of reproducibility~\cite{song2020optimal,9637803,8885156}. Note that (relay-IoT device) communication channels ($L$) and (TBS-relay) communications channels ($K$) are termed as \textit{relay channels} and \textit{TBS channels} respectively, for ease of presentation in the rest of the paper

\textbf{Scheduling:} At each slot $t$, our considered relayed IoT networks involve two simultaneous scheduling steps -- (i) \textit{sampling} of packets generated at the IoT devices by the relays and (ii) \textit{updating} of sampled packets from the relays to the TBS. The set of IoT devices sampled by relay $n$ is denoted by $S_n(t)$ and total devices sampled by all the relays is $\mathcal{S}(t)=\{\mathcal{S}_1(t),\mathcal{S}_2(t), \dots, \mathcal{S}_N(t)\}$. Similarly, updated devices are denoted as $\mathcal{U}(t)$. When a device $m$ is sampled, it transmits its most recent packet to the serving relay $n_m$ and it will replace any of its older packets at $n_m$ \cite{8885156}. Then when it is updated, the packet is transmitted from relay $n_m$ to the TBS. At any slot, each channel can support at most 1 packet.


This paper focuses on conditions where the relay and TBS channels are lossy, and the IoT devices follow the periodic packet generation model. Note that the scheduler doesn't have any prior information on the periodicity of traffic generation at the IoT devices. For scenarios under ideal channel conditions where the wireless channels are lossless and the IoT devices employ generate-at-will traffic generation model, it has been proven in~\cite{9637803} that the MAF-MAD is the optimal AoI scheduler for minimizing the AoI at the TBS, and hence such scenarios will not be considered in this study.

\subsection{Age of Information (AoI) at the relay and TBS} We employ Age of Information (AoI) to measure the freshness of information. In particular, AoI is defined as the \textit{time elapsed since the generation time of the most recent packet received (by the relay/TBS in our case).} 

The AoI of IoT device $m$ at relay $n_m$ is denoted as 

\begin{align}
    AoI_m^{relay} (t) = t - \tau_{s,m} 
    \label{eqn:AoI_evolution_relay}
\end{align}

\noindent where $t$ is the current slot and $\tau_{s,m}$ is the generation time of the \textit{most recently sampled} packet of device $m$ that was received successfully at relay $n_m$. Similarly, the AoI of device $m$ at the TBS is 
\begin{align}
    AoI_m^{TBS} (t)  = t - \tau_{u,m}
    \label{eqn:AoI_evolution_TBS}
\end{align}

\noindent where $\tau_{u,m}$ is the generation time of the \textit{most recently updated} packet of the $m$th IoT device that was received successfully at the TBS. Based on Eqn.~\eqref{eqn:AoI_evolution_relay} and \eqref{eqn:AoI_evolution_TBS}, AoI increases linearly in slots of no reception and decreases at reception instants.

The evolution of AoI with time at the relays and TBS is described next. At $t$, if device $m$ was selected for sampling, i.e., $m \in \mathcal{S}(t)$, its AoI at the relay changes as 

\begin{equation} \label{eqn:transition_AoI_relay_sample}
    AoI_{m}^{relay}(t+1) =
    \begin{cases}
    t+1-\tau_{s,m} & \text{with prob.} \hspace{.05in} 1-l_{s,m}  \\
    AoI_{m}^{relay}(t)+1 & \text{with prob.} \hspace{0.05in} l_{s,m} 
    \end{cases}
\end{equation}

Else if the device $m$ was not sampled, i.e., $m \notin \mathcal{S}(t)$, its AoI at the relay changes as 

\begin{equation} \label{eqn:transition_AoI_relay_no_sample}
    AoI_{m}^{relay}(t+1) = AoI_{m}^{relay}(t)+1
\end{equation}

Similarly AoI for IoT device $m$ at the TBS when selected for update such that $m \in \mathcal{U}(t)$ and when not selected for update such that $m \notin \mathcal{U}(t)$ are shown in Eqn. \eqref{eqn:transition_AoI_TBS_update} and  \eqref{eqn:transition_AoI_TBS_no_update}.

\begin{equation} \label{eqn:transition_AoI_TBS_update}
    AoI_{m}^{TBS}(t+1) = 
    \begin{cases}
     AoI_{m}^{relay}(t)+1 & \text{with prob.} \hspace{.05in} 1-l_{u,m}  \\
    AoI_{m}^{TBS}(t)+1 & \text{with prob.} \hspace{0.05in} l_{u,m} 
    \end{cases}
\end{equation}

\begin{equation} \label{eqn:transition_AoI_TBS_no_update}
    AoI_{m}^{TBS}(t+1) = AoI_{m}^{TBS}(t)+1
\end{equation}

Thus a device's AoI increases at the relays and the TBS if- 
\begin{itemize}
    \item it was sampled/updated but the packet was lost 
    \item it was not sampled/updated.
\end{itemize}

As the information is relayed through the relays, $AoI_{m}^{relay}$ directly impacts $AoI_{m}^{TBS}$. Note that the communication between a device and the TBS has a delay of a single slot, due to which information sampled by a relay at a certain slot cannot be updated to the TBS in the same slot \cite{song2020optimal, 9637803}. 

Finally, the average AoI of all IoT devices at the relays and the TBS during the observation interval $T$ is calculated as Eqn. \eqref{eqn:relay_AoI_avg} and \eqref{eqn:TBS_AoI_avg} respectively

\begin{align}
    AoI^{relay}(T) = \frac{1}{TM} \sum_{t=1}^{T} \sum_{m=1}^{M} AoI_{m}^{relay} (t)
    \label{eqn:relay_AoI_avg}
\end{align}

\begin{align}
    AoI^{TBS}(T) = \frac{1}{TM} \sum_{t=1}^{T} \sum_{m=1}^{M} AoI_{m}^{TBS} (t)
    \label{eqn:TBS_AoI_avg}
\end{align}

\textit{Problem formulation:} Our objective is to design a scheduler that ensures minimum AoI corresponding to all IoT devices at the TBS. Given $S_n(t)$ and $\mathcal{U}(t)$ respectively denote the set of the IoT devices sampled by each relay $n \in N$ and updated to the TBS at a certain slot $t \in T$, we formulate the AoI-aware scheduling problem as the minimization of the average AoI of all IoT devices at the TBS (Eqn. \eqref{eqn:TBS_AoI_avg}) subject to the limited channel constraints, as follows.

 \begin{align}
 	\nonumber & \min_{} AoI^{TBS}(T) \\
 	\text{s.t. } & \left| \mathcal{S}_{n}^{}(t) \right| \le L, \hspace{5pt} \forall n \in N \hspace{5pt} \text{and} \hspace{5pt} t=1,2...T \label{constraint1}
 	\\
 	& \left| \mathcal{U}^{}(t) \right| \le K, \hspace{5pt} t=1,2...T \label{constraint2}
 \end{align}
 
\noindent where constraint \eqref{constraint1} and \eqref{constraint2} refer to the limited number of channels for the sampling and updating respectively. The notations used in this paper are summarized in Table \ref{tab:notations}.

\begin{table}
 \centering
 \small
 \vspace{-0.18in}
 \caption{Notations}
    \begin{tabular}{|p{5.9cm}|p{2cm}|}
    \hline
    \textbf{Meaning }  & \textbf{Symbol} \\ \hline
    Number of IoT devices  & $M$\\ \hline
    Number of relays  & $N$\\ \hline
    IoT devices associated to relay $n$  & $m_n$ \\ \hline
    relay providing coverage to device $m$ & $n_m$ \\ \hline
    Packet generation periodicity of device $m$ & $p_m$  \\ \hline
    AoI of device $m$ at relay & $AoI_m^{relay}$ \\ \hline
    AoI of device $m$ at TBS & $AoI_m^{TBS}$ \\ \hline
    Packet loss between device $m$ and relay $n_m$ & $l_{s,m}$ \\ \hline
    Packet loss between device $m$ and TBS & $l_{u,m}$ \\ \hline
    Devices sampled at $t$ by relay $n$  & $\mathcal{S}_n(t) $ \\ \hline
    Devices sampled at $t$=$\{\mathcal{S}_1(t), \mathcal{S}_2(t),..\mathcal{S}_N(t)\} $  & $\mathcal{S}(t) $ \\ \hline
    Devices updated at $t$  & $\mathcal{U}(t) $  \\ \hline
    Channels at each relay & $L$ \\ \hline
    Channels at the TBS & $K$ \\ \hline
 \end{tabular}
 \label{tab:notations}
\end{table}

\section{Proposed v-PPO based AoI scheduler}
\label{sec:ppo_policy}

This section presents the details of the proposed voting mechanism based proximal policy optimization (v-PPO) based AoI scheduler for two-hop IoT networks with relays. 
The proposed v-PPO based solution is better than the commonly used Deep Q Networks (DQN) for AoI minimization used in the literature~\cite{9014214, zhou2019deep, hu20201, abd2019deep, song2020optimal}.  While DQN can learn the channel qualities and traffic generation patterns,  it has two major disadvantages. First, 
DQN cannot perform well when the action space is large, which is usually the case in larger IoT networks~\cite{dulac2015deep}. Secondly, DQN significantly reduce exploring~\cite{mnih2013playing} its environment after a pre-specified time; thus it cannot adapt to changed network conditions. Details on the proposed v-PPO based AoI scheduler and a comparison between the v-PPO and DQN schedulers are discussed below.

\subsubsection{PPO Preliminaries}
\label{sec:prelims}
In this section, we give a brief introduction to deep RL and PPO algorithms.

\paragraph{Episode, Return, and Value function}
In RL, mathematical entities commonly known as agents, learn to make optimal decisions by interacting in an unknown environment and exploiting the received feedback. Usually, Markov Decision Process (MDP) 
is used to simulate such environment that requires decision making in situations where outcome is partly random and partly under the control of the decision maker or agent. MDP is formally defined as the tuple \{$\mathcal{X}$,$\mathcal{A}$,$\mathcal{T}$,$p(x_0)$,$\gamma$\}. Therefore, the environment consists of a transition function $\mathcal{T}:\mathcal{X}\times\mathcal{A}\rightarrow p(\mathcal{X})$ and a reward function $\mathcal{R}:\mathcal{X}\times\mathcal{A}\times\mathcal{X}\rightarrow \mathbb{R}$. At slot $t$, the agent observe some state $x_t\in \mathcal{X}$ and picks an action $a_t\in \mathcal{A}$ with policy $\pi(a_t|x_t):\mathcal{X}\rightarrow p(\mathcal{A})$. As a result, the environment transitions to a next state $x_{t+1}\sim \mathcal{T}(\cdot|x_t,a_t)$ and returns a scalar reward $r_t=\mathcal{R}(x_t,a_t,x_{t+1})$. The first state $x_0$ is sampled from the initial state distribution $p(x_0)$. Finally, $\gamma \in [0,1]$ is the discount factor.

The agent interacts with the environment until it reaches the terminal state or reaches the time limit $T$, completing an episode $\tau$=\{$x_0$,$a_0$,$r_0$,$\dots$,$x_t$,$a_t$,$r_t$,$\dots$,$x_{T-1}$,$a_{T-1}$,$r_{T-1}$,$x_{T}$\}\footnote{Observation interval $T$ is much smaller than simulation duration $\mathbb{T}$.}. Thus quality of state $x_t$ can be realized as the cumulative sum of rewards from $x_t$ to the end of the episode $\tau$, which is known as the return $G_t$, see Eqn.~\eqref{return_eq}. However, due to the stochastic nature of the environment, there can be many possible next states $x_{t+1}$, resulting multiple return $G_t$ values from state $x_t$ for different episodes. Therefore, quality of state $x_t$ is defined by the value function $V(x_t)$, as the expected return from state $x_t$, see Eqn.~\eqref{value_eq}.
\begin{gather}
    \label{return_eq}
    G_t=\sum_{\hat{t}=t}^{T} \gamma^{\hat{t}-t} r_{\hat{t}}\\
    \label{value_eq}
    V(x_t)=\mathbb{E}\left[ G_t \mid x_t\right]
\end{gather}

\paragraph{Actor-Critic methods}
The primary objective of the agent is to behave optimally with policy $\pi(a_t|x_t)$ so that it can obtain a high cumulative reward $R_{\tau}=r_0+r_1+r_2+\dots+r_{T-1}$ in episode $\tau$. In stochastic policy gradient methods, we maximize the cost function $J(\pi)=\sum_{\tau=1}^{\infty} p(\tau) R_{\tau}$ with respect to the policy $\pi(a_t|x_t)$ to increase the likelihood of sampling the most rewarding episode, where $p(\tau)$ is the probability of sampling episode $\tau$ with current policy $\pi(a_t|x_t)$.

In practice, the state $x_t$ can be multi-dimensional, making it infeasible for traditional table based RL approaches~\cite{watkins1992q}. This provides a clear path for the application of universal function approximators (i.e, neural networks) to model policy $\pi(a_t|x_t)$ with parameter $\theta$. Similarly, value function $V(x_t)$ is modeled with parameters $\psi$. The parameterized policy $\pi(a_t|x_t;\theta)$ and the value function $V(x_t;\psi)$ is commonly known as the \textbf{actor} and the \textbf{critic} networks respectively.

While the actor network interacts with the environment and collect experiences \{$x_t$,$a_t$,$r_t$,$x_{t+1}$\}, the advantage is computed as shown in Eqn.~\eqref{advantage_eq}. Advantage $A_t$ represents the incentive obtained by taking action $a_t$ at state $x_t$ from the prospective of the critic network. Subsequently, the actor and the critic network is jointly optimized by back-propagating the gradients from Eqn.~\eqref{actor_grad} and Eqn.~\eqref{critic_grad}.
\begin{gather}
    \label{advantage_eq}
    A_t=r_t+\gamma V(x_{t+1};\psi)-V(x_t;\psi)\\
    \label{actor_grad}
    \nabla_{\theta}J(\theta)=\mathbb{E}\left[\nabla_{\theta}\log{\pi(a_t|x_t;\theta)}A_t\right] \\
    \label{critic_grad}
    \nabla_{\psi}\mathcal{L}(\psi)=\mathbb{E}\left[(G_t-V(x_t;\psi))\nabla_{\psi}V(x_t;\psi)\right]
\end{gather}

\paragraph{Proximal Policy Optimization}
The actor-critic algorithms that rely on the policy gradient formulation suffer from divergent behavior unless we impose a mechanism to constrain the change of the actor's policy. The primary reason is the on-policy data collection for optimizing the actor and the critic network. This issue is addressed in \cite{schulman2015trust} by introducing trust-region policy updates. Since then, actor-critic algorithms have excelled at tackling real-world problems due to their increased efficiency and stability with respect to Q-Learning. Proximal Policy Optimization (PPO) is the recent addition~\cite{schulman2017proximal} to such class of algorithms and therefore inherits the aforementioned properties. Moreover, PPO improves the efficiency further and achieves trust-region policy updates by simply clipping the probability ratio $\zeta_t\in [1-\epsilon,1+\epsilon]$ between the current $\pi(a_t|x_t;\theta)$ and the old $\pi(a_t|x_t;\theta_{old})$ actor policies. PPO optimizes the actor-network with the gradients shown in Eqn.~\eqref{ppo_grad} and also leverages entropy $\mathcal{H}(\theta)$ based regularization to encourage exploration at the early stages of learning. The \textbf{combined gradient} in Eqn.~\eqref{combined_ppo_grad} optimizes the \textit{actor} and the \textit{critic} networks simultaneously, where $c_1$, $c_2$ are constants.
\begin{gather}
    \label{ppo_grad}
    \nabla_{\theta}J(\theta)=\mathbb{E}\left[\nabla_{\theta}Min(\zeta_t A_t,\left[\zeta_t\right]_{1-\epsilon}^{1+\epsilon} A_t)\right]\\
    \text{where,}~~\zeta_t=\frac{\pi(a_t|x_t;\theta)}{\pi(a_t|x_t;\theta_{old})}\notag \\
    \label{combined_ppo_grad}
    \nabla_{\theta\cup\psi}\mathcal{L}(\theta\cup\psi)=-\nabla_{\theta}J(\theta)+c_1\nabla_{\psi}\mathcal{L}(\psi)-c_2\nabla_{\theta}\mathcal{H}(\theta)
\end{gather}

\subsubsection{MDP formulation}
\label{subsec:MDP_form}
In this section, we formulate the packet scheduling to minimize AoI at relay-relayed IoT networks as an MDP. The state, action and reward in the context of our work is explained next. 

\paragraph{State Space}
State is the scenario encountered by the scheduler. At slot $t$, it is defined as
\begin{align}
x_t = \{\hat{t}, \{{AoI_{m}^{relay}(\hat{t})}\}_{m=1}^{M}, \{{AoI_{m}^{TBS}(\hat{t})}\}_{m=1}^{M}\}_{\hat{t}=t}^{t-z+1}
\label{eqn:state_space}
\end{align}

\noindent where $z$ is the stacking size. The stacking size is discussed in Sec.~\ref{subsub:stack_size}. At $t$=1, $AoI_{m}^{relay}(t)$ and $AoI_{m}^{TBS} (t)$ are initialized to 1. The state space is denoted by $\mathcal{X}$.

\paragraph{Action Space}
As described in Sec. \ref{sec:system}, scheduling involves two simultaneous steps of sampling and updating devices at each slot. Therefore the action at each slot $t$ is  
\begin{align}
    a_t = (\mathcal{S}(t), \mathcal{U}(t))
    \label{eqn:action_space}
\end{align}

\noindent where $\mathcal{S}(t)$=\{$S_1(t), S_2(t), ..S_N(t)$\} is the set of all devices sampled by all the relays at slot $t$ such that $|S_n(t)| \le L \hspace{0.05in} \forall \hspace{0.05in} n \in N$. Similarly, $\mathcal{U}(t)$ is the set of devices selected to update their packets to the TBS where $|\mathcal{U}(t)| \le K$. The action space is denoted by $\mathcal{A}$, where $|\mathcal{A}|=(\prod_{i} \Mycomb[m_i]{L}) \times \Mycomb[M]{K}$.

\paragraph{Reward}
As the objective is improving the information freshness at the TBS, the reward is given by the negative of the average AoI of all the devices at the TBS at slot $t+1$
\begin{align}
     r_t=\mathcal{R}(x_t, a_t, x_{t+1})=-\frac{1}{M}\sum_{m=1}^{M} AoI_m^{TBS}(t+1)
    \label{eqn:reward_fn}
\end{align}

Based on action $a_t$, the environment transitions from state $x_t$ to a new state $x_{t+1}$ according to the state transition probabilities described in Eqn.~\eqref{eqn:transition_AoI_relay_sample}, \eqref{eqn:transition_AoI_relay_no_sample}, \eqref{eqn:transition_AoI_TBS_update}, \eqref{eqn:transition_AoI_TBS_no_update} while resulting in a reward $r_t$. Thus it is a finite horizon MDP with finite state and action spaces, which makes it suitable for RL approaches.

\subsubsection{Voting mechanism based PPO (v-PPO) algorithm}
Here, we propose the voting mechanism based PPO algorithm, which efficiently captures the action space $\mathcal{A}$ for sampling and updating packets at relay $n\in N$ and TBS respectively. The advantage of using voting to sample $L$ packets from $m_n$ devices at relay $n \in N$ is following- at slot $t$, we compute votes $v_{i}^{t} \hspace{0.05in} \forall \hspace{0.05in} i \in m_n$ devices and sample packets from $S_{n}(t)$ devices with top $L$ votes as shown in Eqn. \eqref{vote_sample}. Hence, we can capture $\Mycomb[m_n]{L}$ possible sampling actions with $m_n$ votes. Similarly, we can update $K$ packets to the TBS from $M$ devices with Eqn.~\eqref{vote_update}. The benefit of this approach will be clearer if we consider all $N$ relays and the TBS, since we only have to compute $(\sum_{n=1}^{N} m_n) +M=2M$ votes to represent an action space with cardinality $|\mathcal{A}|=(\prod_{i} \Mycomb[m_i]{L}) \times \Mycomb[M]{K}$.

\begin{gather}
    \label{vote_sample}
    \mathcal{S}_{n}(t)=\argmax_{\mathcal{S}^{\prime} \subseteq m_{n},|\mathcal{S}^{\prime}|\le L} \{v_{i}^{t}\}_{i=1}^{m_n} \hspace{0.05in} \forall \hspace{0.05in} n \in N\\
    \label{vote_update}
    \mathcal{U}(t)=\argmax_{\mathcal{U}^{\prime} \subseteq M,|\mathcal{U}^{\prime}|\le K} \{v_{i}^{t}\}_{i=1}^{M}
\end{gather}

\begin{figure}
    \centering
    \includegraphics[width=.8\columnwidth]{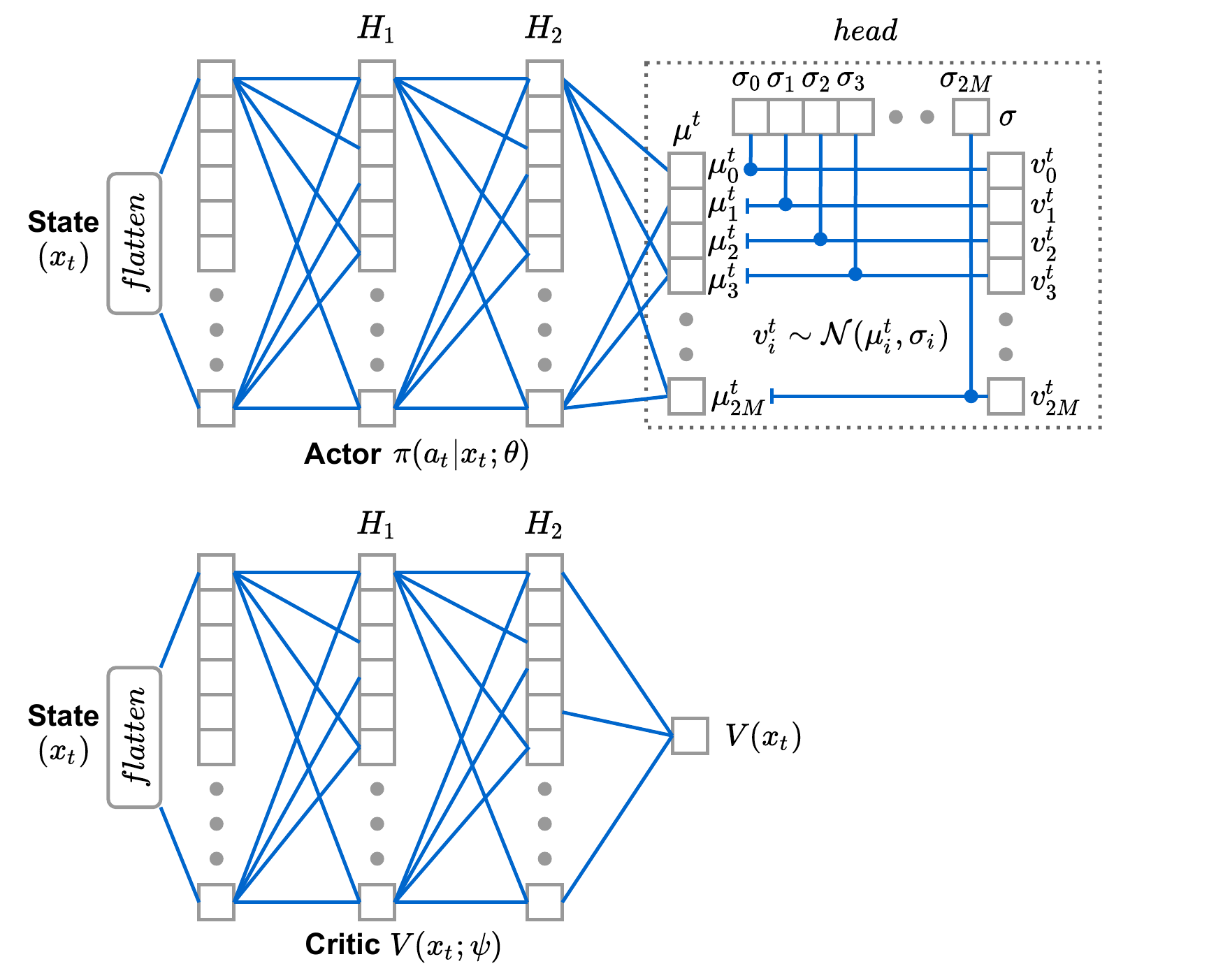}
    \caption{Actor and Critic network architectures}
    \label{fig:ac_net}
    \vspace{-0.05in}
\end{figure}

The proposed AoI scheduler leverages this voting mechanism based PPO algorithm, as shown in Fig. \ref{fig:ac_net}. At slot $t$, the actor network $\pi(a_t|x_t;\theta)$ outputs $2M$ normal distributions $\{\mathcal{N}(\mu_{i}^{t},\sigma_{i})\}_{i=1}^{2M}$ with mean $\mu^{t}=\{\mu_{i}^{t}\}_{i=1}^{2M}$ and standard deviation\footnote{Note that, PPO based scheduler explores the action space by increasing entropy $\mathcal{H}(\theta)=\frac{1}{2}\log(2\pi\sigma^2)+\frac{1}{2}$, see Eqn.~\eqref{combined_ppo_grad}.} $\sigma=\{\sigma_{i}\}_{i=1}^{2M}$ to compute the votes as $\{v_{i}^{t}:v_{i}^{t}\sim \mathcal{N}(\mu_{i}^{t},\sigma_{i})\}_{i=1}^{2M}$. Thereafter, the votes are converted to $\mathcal{S}_n(t) \hspace{0.05in} \forall \hspace{0.05in} n \in N$ and $\mathcal{U}(t)$ as per Eqns.~\eqref{vote_sample} and~\eqref{vote_update}, which are combined to obtain the final action $a_t$ as given in Eqn.~\eqref{eqn:action_space}.

\subsubsection{Algorithmic Description}
\begin{algorithm}
\caption{v-PPO based AoI scheduler}
\label{algo:ppo_scheduler}
\begin{algorithmic}[1]
\State Initialize actor $\pi(a_t|x_t;\theta)$, critic $V(x_t;\psi)$
\State Initialize actor's old parameters, $\theta_{old} \gets \theta$
\For{iteration $\gets 1,2$ \dots $\lceil\frac{\mathbb{T}}{\mathcal{E}\times T}\rceil$}
    \State Replay Buffer $\mathcal{B}\gets \{\}$
    \For{episode $\gets 1,2$ \dots $\mathcal{E}$}
        \For{slot t $\gets 0,1,2, \dots T-1$}
            \State Observe current state $x_t$
            \State Execute action $a_t\sim\pi(\cdot|x_t;\theta_{old})$
            \State Observe next state $x_{t+1}$ and, reward $r_t$
            \State Compute Advantage $A_t$ using Eqn.~\eqref{advantage_eq}
            \State $\mathcal{B}\gets \mathcal{B}\cup \{x_t,a_t,r_t,x_{t+1},A_t\}$
        \EndFor
    \EndFor
    \For{$epoch \gets 1,2,\dots \eta$}
        \State Sample minibatch $b_{m}\sim \mathcal{B}$ of size $|b_{m}|\leq E\times T$
        \State Compute gradients $\nabla_{\theta\cup\psi}\mathcal{L}(\theta \cup \psi)$ using Eqn.~\eqref{combined_ppo_grad}
        \State Optimize parameters $\theta$ and $\psi$ with the gradients:
        \State\indent $\theta \gets \theta -\alpha \nabla_{\theta}\mathcal{L}(\theta \cup \psi)$
        \State\indent $\psi \gets \psi -\alpha \nabla_{\psi}\mathcal{L}(\theta \cup \psi)$
    \EndFor
    \State Update actor's old parameters, $\theta_{old} \gets \theta$
\EndFor
\end{algorithmic}
\end{algorithm}

As shown in Algorithm \ref{algo:ppo_scheduler}, in the beginning, the v-PPO based AoI scheduler randomly initializes the parameters $\theta$ and $\psi$ of the actor $\pi(.)$ and critic $V(.)$ networks, respectively (Line 1). It also initializes the actor's old parameters $\theta_{old}$ with $\theta$ (Line 2).

Next, the replay buffer $\mathcal{B}$ is (re)initialized to store new experiences (Line 4). Following it, the actor interacts with the environment for $\mathcal{E}$ episodes, where each episode is of length $T$. At slot $t$ of an episode, the actor and the critic network observes the state $x_t$. Note that, the actor network utilizes its old parameters $\theta_{old}$ to compute action $a_t$. Next, action $a_t$ is executed in the environment, resulting transition to next state $x_{t+1}$ and the v-PPO based AoI scheduler receives reward $r_t$(Line 8-9). The critic network computes the value of the state $V(x_t;\psi)$ to estimate the advantage $A_t$ of taking action $a_t$ on state $x_t$ using Eqn.~ \eqref{advantage_eq} (Line 10). At this point, the v-PPO based scheduler obtains a new experience $\{x_t,a_t,r_t,x_{t+1},A_t\}$, which is stored in the replay buffer $\mathcal{B}$ (Line 11). At the end of $\mathcal{E}$ episodes, the replay buffer contains $\mathcal{E}\times T$ entries (Lines 12 - 18), which trains the actor and critic networks as discussed next.

A minibatch $b_m$ is sampled from the replay buffer $\mathcal{B}$, where $|b_m| \leq \mathcal{E}\times T$. Subsequently, the gradient $\nabla_{\theta\cup\psi}\mathcal{L}(\theta \cup \psi)$ is computed on minibatch $b_m$ using Eqn. \eqref{combined_ppo_grad}. Next, the gradient \textit{(i) $\nabla_{\theta}\mathcal{L}(\theta \cup \psi)$ with respect to actor parameters $\theta$, and (ii) $\nabla_{\psi}\mathcal{L}(\theta \cup \psi)$ with respect to critic parameters $\psi$} is utilized for optimizing (gradient descent) $\theta$ and $\psi$, where $\alpha$ is the learning rate. These gradient updates improve the actor and critic parameters and the v-PPO based scheduler performs multiple such updates over $\eta$ epochs on the replay buffer $\mathcal{B}$. At the end of $\eta$ epochs (Line 17), the actor network's old parameters $\theta_{old}$ is updated with the new actor network parameters $\theta$ (Line 18), which will be utilized for interacting with the environment in the next iteration (See line 3). The v-PPO based scheduler runs for $\lceil\frac{\mathbb{T}}{\mathcal{E}\times T}\rceil$ iterations (Line 3-18) that improves the actor network parameters $\theta$ over old parameters $\theta_{old}$, where $\mathbb{T}$ denotes the total simulation duration.

Note that the convergence of deep neural networks are difficult to analyze and strongly depends on the hyper-parameters used \cite{jomaa2019hyp}. Selection of hyperparameters is a challenging task and therefore a reasonable set of hyperparameters is found by trying different values.
Similar to \cite{abd2019deep, 9014214, song2020optimal, zhou2019deep}, we limit investigation of the algorithm's convergence to simulations, where the neural network converges under the hyper-parameters used. The results presented for the DQN and v-PPO based AoI scheduler are the values obtained after their convergence.

\begin{figure}
    \centering
    \includegraphics[width=0.9\columnwidth]{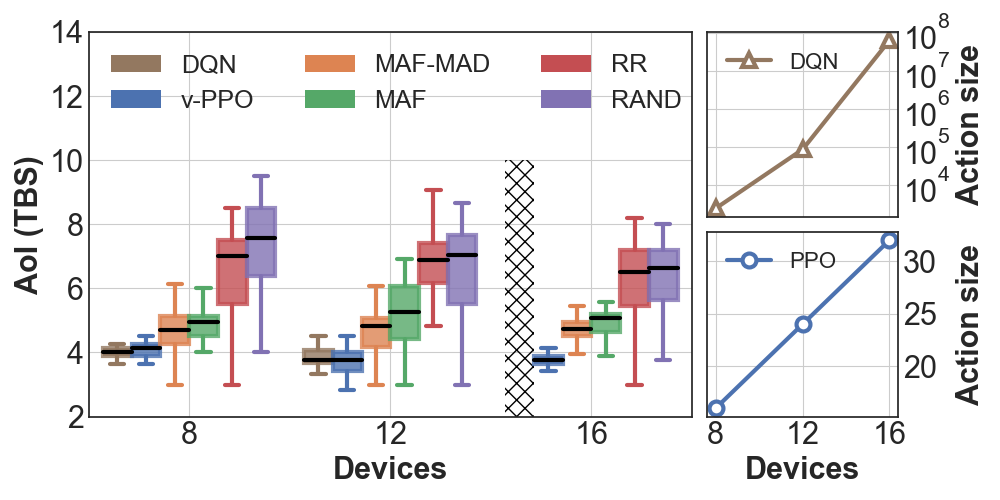}
    \caption{Scalability of DQN vs v-PPO based AoI scheduler.}
    \label{fig:dqn_vs_ppo_scalability}
    \vspace{-0.05in}
\end{figure}

\subsection{v-PPO based AoI scheduler vs DQN-based AoI scheduler}
\label{subsec:ppo_vs_dqn}



This subsection highlights the key limitations of existing DQN based AoI scheduler, and discusses how the proposed v-PPO based AoI scheduler addresses them.

\paragraph{Smaller Action Size} DQN suffers from performance issues in networks with larger action spaces~\cite{9637803}. The number of actions for the DQN-based scheduler is given by $ (\prod_{i} \Mycomb[m_i]{L}) \times \Mycomb[M]{K}$, which means the action size increases rapidly as the network size increases and DQN \textit{doesn't} provide satisfactory performance when the action size is very large \cite{zahavy2018learn}. Due to this, works involving DQN for scheduling in IoT networks to improve AoI have been limited to smaller networks~\cite{9014214, zhou2019deep, hu20201, abd2019deep, song2020optimal}. This can be seen in Fig.~\ref{fig:dqn_vs_ppo_scalability} where we consider 3 different IoT networks with relays with $M = 8, 12,$ and $16$ IoT devices. Note that the number of relay channels ($L$) is $\frac{m_n}{2} $ and TBS channels ($K$) is $\frac{M}{2}$ as it leads to the maximum possible actions (see Sec.\ref{subsec:MDP_form}). 

As shown in Fig.~\ref{fig:dqn_vs_ppo_scalability}, the DQN-based scheduler sees a factorial increase in the action size, whereas the action size of the v-PPO based scheduler increases linearly as $2M$, twice the number of IoT devices $M$. While both DQN and v-PPO perform relatively well for the first two networks with $M=8$ and $12$ devices, the DQN-based approach is not feasible for $M=16$ as its action size is of the order $10^8$ (memory required is very large), whereas the v-PPO's action size is $32$. This shows several folds improvement of v-PPO based scheduler over DQN-based scheduler in reducing the action size, and thus, much better scalability of the v-PPO based AoI scheduler.

\paragraph{Scheduler's Generalization ability}
\label{label:mod_gen}
In certain situations, the training and the deployment network conditions could be different, e.g., devices might change the periodicities or the channel qualities can change. Under such situations, it is desirable if a trained scheduler is able to adapt to the new environment faster than training a new scheduler from scratch. Because v-PPO is an on-policy algorithm, it is able to adapt. Whereas DQN is an off-policy algorithm that will always need a new scheduler to be trained every time there is a change in the network conditions. 

\begin{figure}[htb]
\centering
\subfigure[\label{fig:channel_adapt}]{
\epsfig{figure=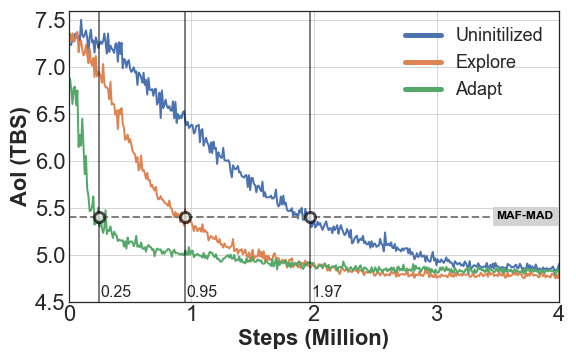,width=0.8\columnwidth , keepaspectratio}}
\subfigure[\label{fig:period_adapt}]{
\epsfig{figure=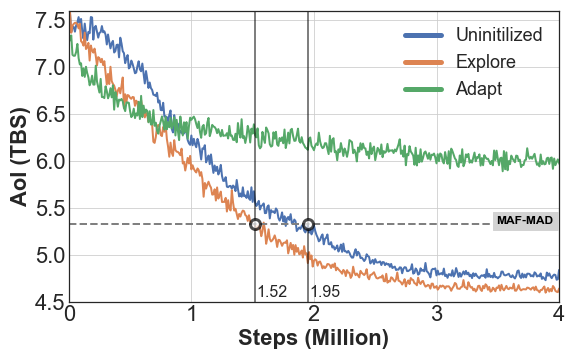,width=0.8\columnwidth , keepaspectratio}}
\caption{Adaptability of the uninitialized, explore and adapt v-PPO schedulers with respect to changed (a) channel conditions (b) device periodicities. Here uninitialized refers to no pre-training, whereas the explore and adapt refers to pre-trained models with and without the actor head being retrained.}
\vspace{-0.05in}
\end{figure}



Consider a IoT network with relay $E_1$ with the parameters given in~Table. \ref{tab:network_e1}. Let us take two different networks, $E_2$ and $E_3$ both of which are similar to $E_1$ with the only difference mentioned in the following-

\begin{itemize}
    \item In network $E_2$, the channel conditions for $10\%$ IoT devices and relays have changed with respect to $E_1$,
    \item In network $E_3$ the periodicity for $10\%$ of IoT devices have changed with respect to $E_1$.
\end{itemize}

In both $E_2$ and $E_3$, a v-PPO based AoI scheduler will be trained from scratch to get the baseline performance and this will be termed as \textit{uninitialized}. To showcase the ability of the v-PPO based AoI scheduler to generalize to a changed network environment, a v-PPO based scheduler trained in $E_1$ will be deployed in $E_2$ and $E_3$ using 2 different approaches -- 

(i) \textit{explore}, meaning that the scheduler trained for $E_1$ will be reused except for the actor head which will be re-trained;

(ii) \textit{adapt}, meaning that scheduler trained for $E_1$ will be entirely reused including the actor head. 

\noindent Note that, the actor head is comprised of the output layer of the actor-network as shown in Fig. \ref{fig:ac_net}.

Fig.~\ref{fig:channel_adapt} shows the results for network $E_2$. Both \textit{explore} and \textit{adapt} scheduler converge earlier than the \textit{uninitialized}, with the \textit{adapt} scheduler converging exceptionally fast. Therefore additional exploration to retrain the actor head doesn't lead to any benefit. However Fig.~\ref{fig:period_adapt} shows the results for network $E_3$ where the \textit{adapt} scheduler is unable to perform well and doesn't even outperform the MAF-MAD scheduler. Therefore re-training the actor head becomes important. The performance of the \textit{uninitialized} and \textit{adapt} schedulers are similar, signifying that the v-PPO based scheduler generalizes  well for changed channel conditions as compared to changed device periodicity.

\section{Performance Evaluation}
\label{sec:results}

This section discusses the performance evaluation of the proposed v-PPO based AoI scheduler against the baseline traditional and DQN-based AoI schedulers in terms of minimizing AoI at the TBS. The IoT devices and relays are placed randomly \cite{lyu2016placement} in a simulation area of $l= b = 1000$ m and the observation interval $T = 20$ slots. Unless otherwise stated,  we consider an IoT network setting with $30$ IoT devices which are uniformly distributed among the $3$ relays. The default parameters for the simulation study are listed in Table~\ref{tab:network_e1}. As discussed in the previous section, because of the large memory requirements, it is not feasible for the DQN-based scheduler to deal with a related large IoT network simulation setting (with $> 12$ IoT devices), we do not showcase the results for the DQN-based scheduler. 

The v-PPO based AoI scheduler (and DQN-based AoI scheduler results presented in the previous section \ref{subsec:ppo_vs_dqn}) are implemented on an NVIDIA DGX station with an Intel Xeon E5-2698 v4 CPU and an NVIDIA Tesla V100 GPU (32 GB memory). Data communication between CPU and GPU takes place through a PCIe 3.0 X16 slot. The hyper-parameters for DQN are kept the same as in~\cite{9637803}, while the hyper-parameters for the v-PPO implementation are shown in Table~\ref{tab:hyperparameters_PPO} respectively.

\begin{table}[htb]
\centering
\small
\vspace{-0.1in}
\caption{Default network parameters}
\begin{tabular}{|c|c|}
\hline
Number of IoT devices, $M$ & $30$                                                \\ \hline
Number of relays, $N$  & $3$                                                \\ \hline
IoT devices per relay, $m_n$  & $(10,10,10)$                                               \\ \hline
Channels at each relay, $L$            & $4$                                                \\ \hline

Channels at the TBS, $K$            & $10$                                                \\ \hline

\end{tabular}
\label{tab:network_e1}
\end{table}

\begin{table}[htb]
\centering
\small
\vspace{-0.1in}
\caption{PPO Hyper-parameters}
\begin{tabular}{|c|c|}
\hline
Neurons in hidden layer 1 $H_1$   & $256$                                                \\ \hline
Neurons in hidden layer 2 $H_1$   & $256$                                                \\ \hline
Relay memory size $|\mathcal{B}|$ & $2048$                                               \\ \hline
Minibatch size $|b_m|$            & $512$                                                \\ \hline
Learning rate $\alpha$            & $2.5\times10^{-4}$ 
     \\ \hline
Discount factor $\gamma$          & $0.99$                                               \\ \hline
Activation function               & ReLU                                                 \\ \hline
Clip range $\epsilon$            & $0.2$                                                \\ \hline
Optimizer                         & Adam                                                 \\ \hline
Epochs $\eta$                     & $10$                                                 \\ \hline
Total episodes $\mathcal{E}$      & $5\times 10^{5}$                                     \\ \hline
\end{tabular}
\label{tab:hyperparameters_PPO}
\end{table}

\subsection{Baseline schedulers} \label{sec:baseline_schedulers}

\begin{itemize}

    \item \textbf{Maximal AoI First - Maximal Age Difference (MAF-MAD) scheduler} -- this scheduler was proposed in~\cite{9637803} where devices with the highest AoI at the relay are selected for sampling, and devices with highest AoI difference between the relay and the TBS ($AoI_m^{diff} (t)$ ) are selected for updating. $AoI_m^{diff}(t)$ for device $m$ is calculated as -
    
    \begin{align}
        AoI_m^{diff} (t) = AoI_{m}^{TBS}(t) - AoI_{m}^{relay}(t) 
        \label{eqn:AoI_users_diff_update}
    \end{align}
 
    \item \textbf{Maximal AoI First (MAF) scheduler} -- devices with the highest AoI at both the relay and TBS are selected for sampling and updating respectively \cite{sun2018age}. For sampling, each relay selects $L$ devices with the highest $AoI_m^{relay}(t)$ out of the $m_n$ devices associated to it. Similarly for updating, $K$ devices out of the total $M$ with highest $AoI_m^{TBS} (t)$ update their packets.
    
    \item \textbf{Deep-Q Network (DQN)-based scheduler} -- this was proposed in a recent work~\cite{9637803} and is based on a centralized DQN agent at the TBS. 
    
    \item \textbf{Round Robin (RR) Scheduler} -- under RR, the available channels are assigned in an equal and circular fashion among the devices which ensures fairness among the devices \cite{muller2018evaluation}
    
    \item \textbf{Random scheduler} -- A random set of devices are selected to be sampled and updated. 

\end{itemize}

\begin{figure}
    \centering
    \includegraphics[width=0.8\columnwidth]{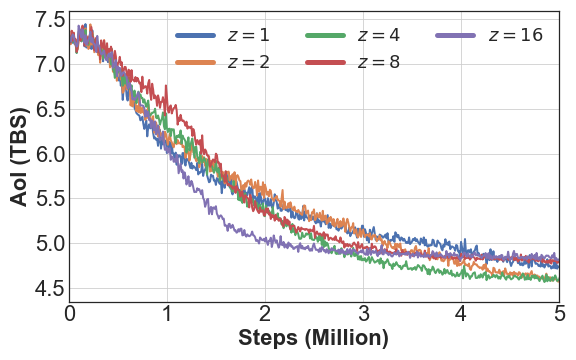}
    \caption{Stack Size ($z$)}
    \label{fig:stack_size}
    \vspace{-0.05in}
\end{figure}

\subsection{AoI Stacking}
\label{subsub:stack_size}
As shown in Eqn.~\eqref{eqn:state_space}, the state space consists of the AoI of the devices stacked for previous $z$ slots. A higher stack size provides more historical information at the expense of larger memory requirements. Here, Fig.~\ref{fig:stack_size} shows the AoI at the TBS achieved for different stack sizes. The convergence is initially faster with $z=16$, but it takes more time to reach the optimal value. Whereas $z=1$ converges very slowly. As, with higher stack size and increased information in the state, the parameters of the v-PPO based scheduler increase as well. Hence it takes less time to reach a considerable performance but struggles to fine-tune itself against the environment. $z=4$ provides a good trade-off between the convergence speed and the optimal performance. Thus the remaining experiments will utilize $z=4$ slots for the state space. 

\subsection{Ideal vs practical environment}
\begin{figure}
    \centering
    \includegraphics[width=0.8\columnwidth]{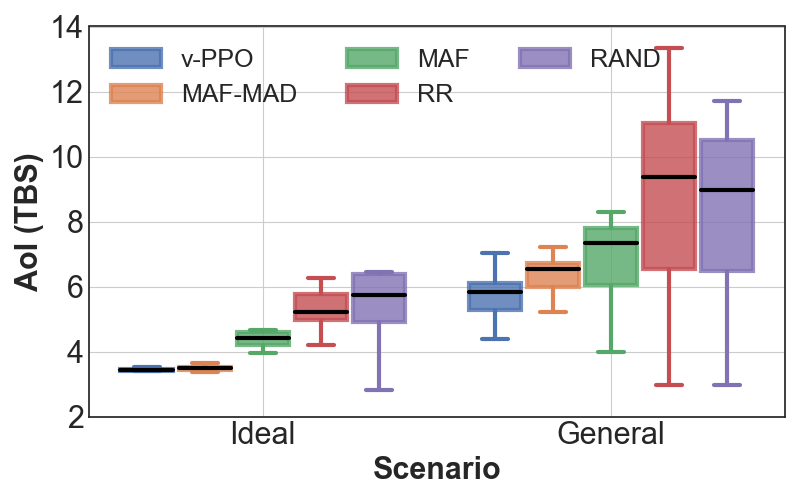}
    \caption{Ideal vs practical IoT network environment}
    \label{fig:ideal_vs_nonideal}
    \vspace{-0.05in}
\end{figure}

As mentioned before, ideal environments are characterized by lossless channels and a generate-at-will traffic generation policy at the devices. The results shown in Fig. \ref{fig:ideal_vs_nonideal} confirm the analytical results derived in~\cite{9637803} that MAF-MAD is the optimal scheduling policy and outperforms all other scheduling policies under such ideal environments. As our proposed v-PPO based scheduler is also able to learn the optimal scheduling, it’s performance also converges to MAF-MAD’s performance. Overall, the v-PPO and MAF-MAD outperform MAF, RR, random by $22.3\%$, $42.1\%$, $38.3\%$ respectively. However, when a practical environment is considered where the ideal assumptions do not hold, MAF-MAD is no longer the optimal scheduler and v-PPO does better. It outperforms MAF-MAD, MAF, RR, random by $10.6\%$, $25.1\%$, $30.9\%$, $37.9\%$ respectively. While the results could be shown for any channel condition and periodicities, they are pre-assigned for ease of reproducibility \cite{song2020optimal,9637803,8885156}.

Note that the range of the AoI achieved by the v-PPO based AoI scheduler is much less than the other schedulers. The v-PPO based scheduler considers the channel conditions while allocating resources. Therefore the devices with bad channels are scheduled more frequently so that their AoI do not suffer, at the expense of a higher AoI of the devices with better channels. However, for the other schedulers, the devices with good channels see a very low AoI, and devices with bad channels will have high AoI, resulting in a higher range. This is because, unlike our v-PPO based scheduler, the counterpart schedulers do not have the ability to learn the various channel conditions/fluctuations.

For the rest of the experiments below, real-world, practical network conditions with lossy channels and periodic packet generation at the IoT devices are considered. 

\begin{figure}
    \centering
    \includegraphics[width=0.8\columnwidth]{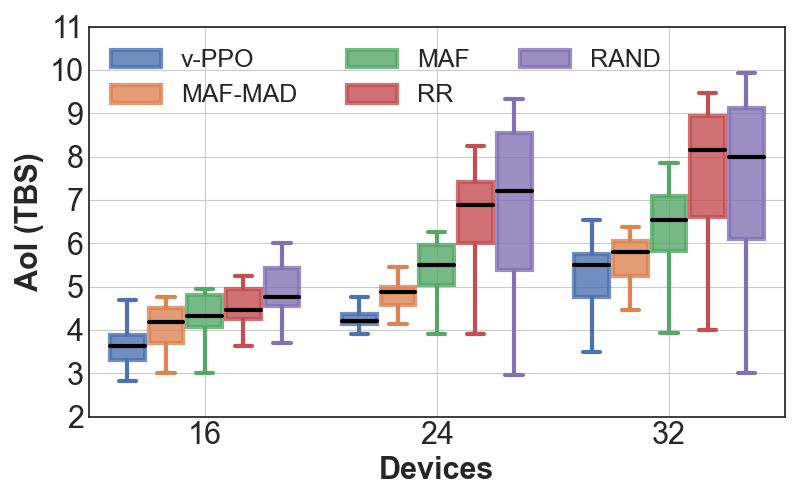}
    \caption{AoI for varying number of IoT devices ($M$)}
    \label{fig:increase_users}
    \vspace{-0.05in}
\end{figure}

\subsection{Simulation results}
In this section, we analyze the effectiveness of the v-PPO based scheduler in minimizing the AoI at the TBS under practical scenarios; for varying parameters such as, number of IoT devices, relays, relay channels, and TBS channels.

\paragraph{Varying number of IoT devices ($M$)}
As shown in Fig. \ref{fig:increase_users}, the v-PPO based AoI scheduler outperforms the baselines in all considered scenarios by a significant improvement of AoI at the TBS. However, an upward trend in the AoI can be observed as we increase the number of IoT devices in the network. The primary reason is the constant sampling and updating capacity of the network. Therefore, the increase of IoT devices results in less frequent sampling and updating of packets from a particular IoT device. Thus, we observe an increased average AoI at the TBS for all schedulers.

\begin{figure}
    \centering
    \includegraphics[width=0.8\columnwidth]{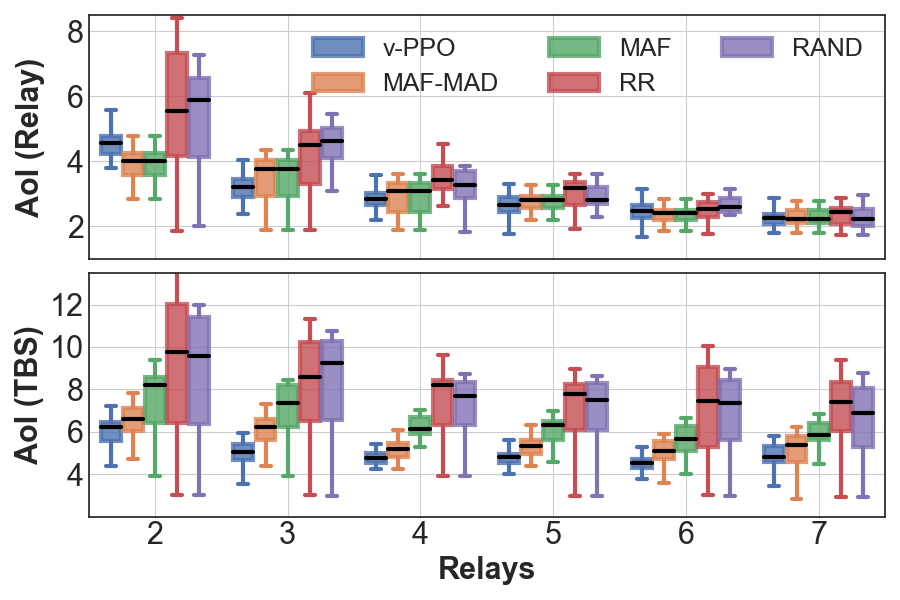}
    \caption{AoI for varying number of relays ($N$)}
    \label{fig:increase_relays}
    \vspace{-0.05in}
\end{figure}

\paragraph{Varying number of relays ($N$)}
The v-PPO based AoI scheduler shows superiority in minimizing AoI at TBS by changing the number of relays in the network, see Fig. \ref{fig:increase_relays}. As shown in the figure, the AoI at relay is improved with more relays, which directly improves the AoI at the TBS as per Eqn \eqref{eqn:transition_AoI_TBS_update}. However, after a point, all schedulers show equivalent AoI at the relay as most IoT devices can be sampled at once, and there is little scope of improvement due to scheduling. Despite similar AoI at the relay, the v-PPO based scheduler outperforms the baselines and improves the AoI further as it employs better scheduling decisions at the TBS.

\begin{figure}
    \centering
    \includegraphics[width=0.8\columnwidth]{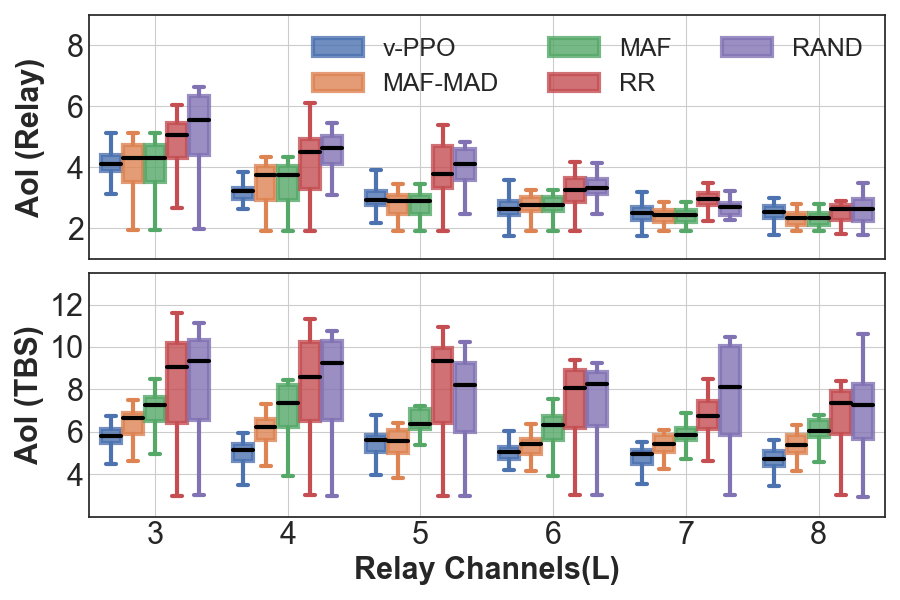}
    \caption{AoI for varying relay Channels ($L$)}
    \label{fig:increase_L}
    \vspace{-0.05in}
\end{figure}

\paragraph{Varying number of relay channels ($L$)}
The increasing number of relay channels has a similar effect to increasing the number of relays. For both ways, the sampling capacity of the network is improved. Hence, the AoI at the relay improves up to a point when most of the IoT devices can be sampled at once and then becomes equivalent for all schedulers, see Fig. \ref{fig:increase_L}. Nevertheless, our proposed v-PPO based scheduler outperforms the baselines in minimizing the AoI at the TBS in all considered scenarios. Note that the v-PPO based scheduler achieves a $61.8\%$ improvement in AoI at the relay with increased relay channels from $3$ to $8$. However, all this gain is not transferable to the TBS, and it only sees a $23.7\%$ improvement in the AoI due to the bottleneck at relay to TBS communication.

\begin{figure}
    \centering
    \includegraphics[width=0.8\columnwidth]{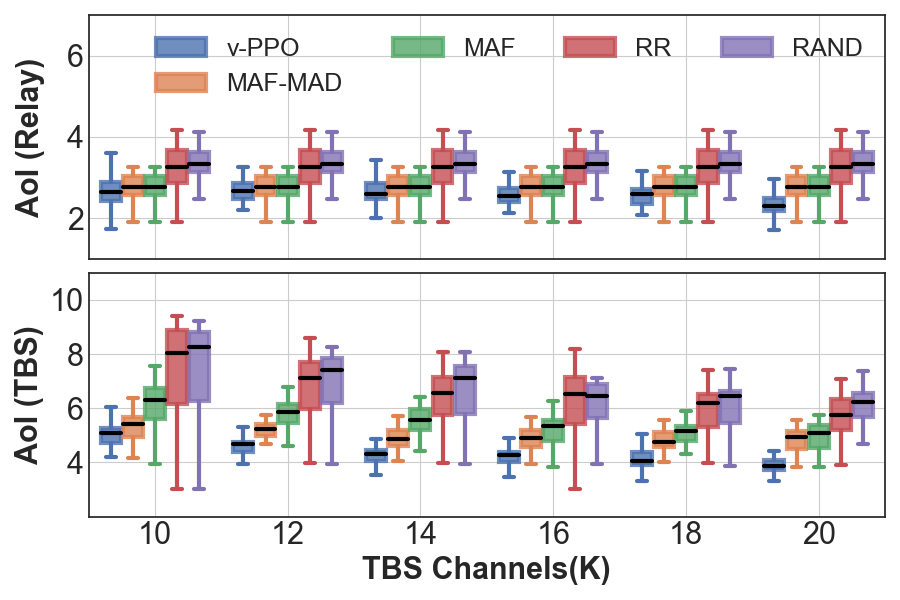}
    \caption{AoI for varying TBS Channels ($K$)}
    \label{fig:increase_K}
    \vspace{-0.05in}
\end{figure}

\paragraph{Varying number of TBS channels ($K$)}
On the contrary, increasing the number of TBS channels does not change the sampling capacity of the network. Thus, the AoI at the relay remains unchanged for a particular scheduler with an increase of $10$ to $20$ in TBS channels, see Fig. \ref{fig:increase_K}. Instead, the network's updating capacity is improved, allowing more packets to be transferred from the relays to the TBS. Therefore, the TBS sees a $31.0\%$ improvement in AoI for v-PPO based scheduling and outperforms the baselines. Thus, the channels between the relays and the TBS ($K$) are more critical bottlenecks than the channels between IoT devices and the relays ($L$) in minimizing AoI at the TBS.

\section{Conclusion}
\label{sec:conclusion}


In this work, we proposed a novel Age of Information (AoI) scheduler for relayed IoT networks, which overcomes the practical challenges of real-world IoT network deployments, including, varying changes in the network conditions, unknown traffic generation patterns at each IoT device, and channel qualities, and scalability to large IoT networks, and, thus, can be practically deployed for improveing information freshness in real-world IoT networks.  Specifically, our proposed AoI scheduler utilizes a voting mechanism based proximal policy optimization (v-PPO) algorithm that maintains a linear action space, making it capable of scaling well with large IoT networks, adapt well to changing network conditions, and learns unknown traffic generation patterns. Our simulation results demonstrate that the proposed AoI scheduler outperforms both traditional (non-ML) and ML-based baseline AoI schedulers in all considered practical scenarios. Our study also noted that the channels between the relay and TBS (i.e., backhaul links) plays a more critical role in minimizing overall AoI at TBS, compared to the channels between IoT devices and relay nodes (i.e., access links).




\ifCLASSOPTIONcaptionsoff
  \newpage
\fi

\bibliographystyle{IEEEtran}
\bibliography{mybib}

\end{document}